\documentclass[conference]{IEEEtran}
\IEEEoverridecommandlockouts
\usepackage{cite}
\usepackage{amsmath,amssymb,amsfonts}
\usepackage{algorithmic}
\usepackage{graphicx}
\usepackage{textcomp}
\usepackage{xcolor}
\usepackage{paralist}
\usepackage{listings}
\usepackage[linesnumbered,ruled,vlined]{algorithm2e}

\SetCommentSty{mycommfont}
\def\BibTeX{{\rm B\kern-.05em{\sc i\kern-.025em b}\kern-.08em
    T\kern-.1667em\lower.7ex\hbox{E}\kern-.125emX}}
\begin{document}

\title{Measuring Stakeholder Agreement and Stability in a Decentralised Organisation\\
}

\author{\IEEEauthorblockN{ Sarad Venugopalan and Heiko Aydt}
	\IEEEauthorblockA{\textit{Singapore-ETH Centre} \\
		%Singapore\\
		sarad.venugopalan@sec.ethz.ch, heiko.aydt@sec.ethz.ch }}

%\and
%\and
%\and
%\and
%\and
%\IEEEauthorblockA{\textit{dept. name of organization (of %Aff.)} \\

\maketitle

\begin{abstract}
	A decentralised organisation (DO) is a  multi-stakeholder institution where decision making is assigned to various levels of the organisation.
	Decentralised stakeholders play an important role in the governance of a decentralised organisation.
	The ability to measure DO stability will help monitor the health of the organisation and acts as an early warning system for disagreement
	and group exit, leading to its destabilisation/collapse. For example, blockchain hard forks.
	We propose the organisational tension quadrilateral to study agreement between stakeholders and build a tool based on voting data (information as vote choices) to measure its stability. The stakeholders  are permitted to vote their choice into an electronic ballot box. Here, each vote choice represents a measure of agreement. When voting ends, this information is aggregated and used to build a metric for DO stability.
	To the best of our knowledge, there are no similar tools available to measure DO stability.
	 
\end{abstract}


\section{Introduction}
\label{sec:intro}
A decentralised organisation~\cite{Codemonk2022} is a  multi-stakeholder institution where decision making is assigned to various levels of the organisation. Each level comprises different individual groups,  with the autonomy to make decisions and  act on it.
A stakeholder may belong to one or more groups in the organisation.
The stakeholders are  tasked with the continued wellbeing of the organisations' value producing system (such as goods and services), governed via cooperation for mutual benefit.
Along with traditional organisations, the DO aims to provide good governance via continuous improvement, and at the same time ensuring organisational stability.
Unlike a decentralised autonomous organisation (DAO) where governance is automated using blockchain smart contracts~\cite{Kiayias2023}, the governance in a DO may not be fully automated.

For context, we use Ethereum governance~\cite{ethgov2023} to see how it resolves disagreements. The mechanism for handling disagreements by Ethereum governance is described as follows ---  “Having many stakeholders with different motivations and beliefs means that disagreements are not uncommon.
Generally, disagreements are handled with long-form discussion in public forums to understand the root of the problem and allow anyone to weigh in. Typically, one group concedes, or a happy medium is achieved. If one group feels strongly enough, forcing through a particular change could result in a chain split. A chain split is when some stakeholders protest implementing a protocol change resulting in different, incompatible versions of the protocol operating, from which two distinct blockchains emerge."
In the time leading to the finalisation of Ethereum Improvement Proposal (EIP)-779 hard fork~\cite{eip779} (as a result of the year 2016 DAO hack~\cite{Siegel2023} on an insecure smart contract running on an Ethereum based DAO), the voting turnout to approve the fork was very low and many people were unaware of the voting.

While such proposals are extensively debated in private and public forums, it remains difficult to quantify stakeholder support and initiate negotiations early on without agreement measurements, and attempt  to arrive at a consensus before a final decision is made.
Our main objective is to run an opinion poll in tandem with important proposals that are under  review.  For e.g., contentious EIPs. The opinion poll is repeated over  fixed intervals in order to measure changing agreement among stakeholders, and to inform public policy-making made by the organisation as part of the process of stable governance. The opinion poll does not change the current governance mechanism. It is used to provide additional information to make informed decisions based on   measurements derived from the poll.
In the context of Ethereum, running the opinion poll vote in tandem with a proposal would  permit stakeholders to signal their support.
%It may also be used to measure the agreement between different Ethereum groups/teams.
It also permits the early detection of disagreement between core teams, potential hard forks,  to gauge support, negotiate,  and make informed decisions.

Building strategic indicator sets typically requires a few simple questions such as --- `What is your agreeability to a given proposal?', `what is your agreeability with a given core groups in the DO?', or `what is your agreeability with exiting the DO?'.
A measurement tool will allow us to understand when a DO   is deemed stable and when it is not. It will assist the organisation to  take  corrective steps and address them in a timely manner.
We make the following contributions,
\begin{compactenum}
	
	\item [\textit{i}.)] We propose a method to study agreement between stakeholders and groups/teams in a DO (see Section~\ref{sec:dostability}).
	
	\item [\textit{ii}.)] We present a voting tool (see Section~\ref{sec:votingtool}) to obtain the information required to quantify stakeholder agreement and measure stability in the DO.

	\item [\textit{iii}.)] We explain our method to compute a stability score and plot time series charts to study key indicator trends (see Section~\ref{sec:experiments}).
\end{compactenum}

The rest of the paper is organised as follows.
Section~\ref{sec:background} provides the background information.
Section~\ref{sec:modelandgoals} presents the system model and design goals.
Section~\ref{sec:dostability} presents the proposed method to study agreement in a DO. Section~\ref{sec:repeatedvoting} discusses the repeated voting framework used.
Section~\ref{sec:votingtool} describes the voting tool used to collect information from the stakeholders.
Section~\ref{sec:experiments} presents the experiments and measurements to assign a stability score for the DO.
We
conclude with our  contribution  in Section~\ref{sec:conclusions}.

\begin{figure*}
	\centering
	\includegraphics[width=1.0\linewidth]{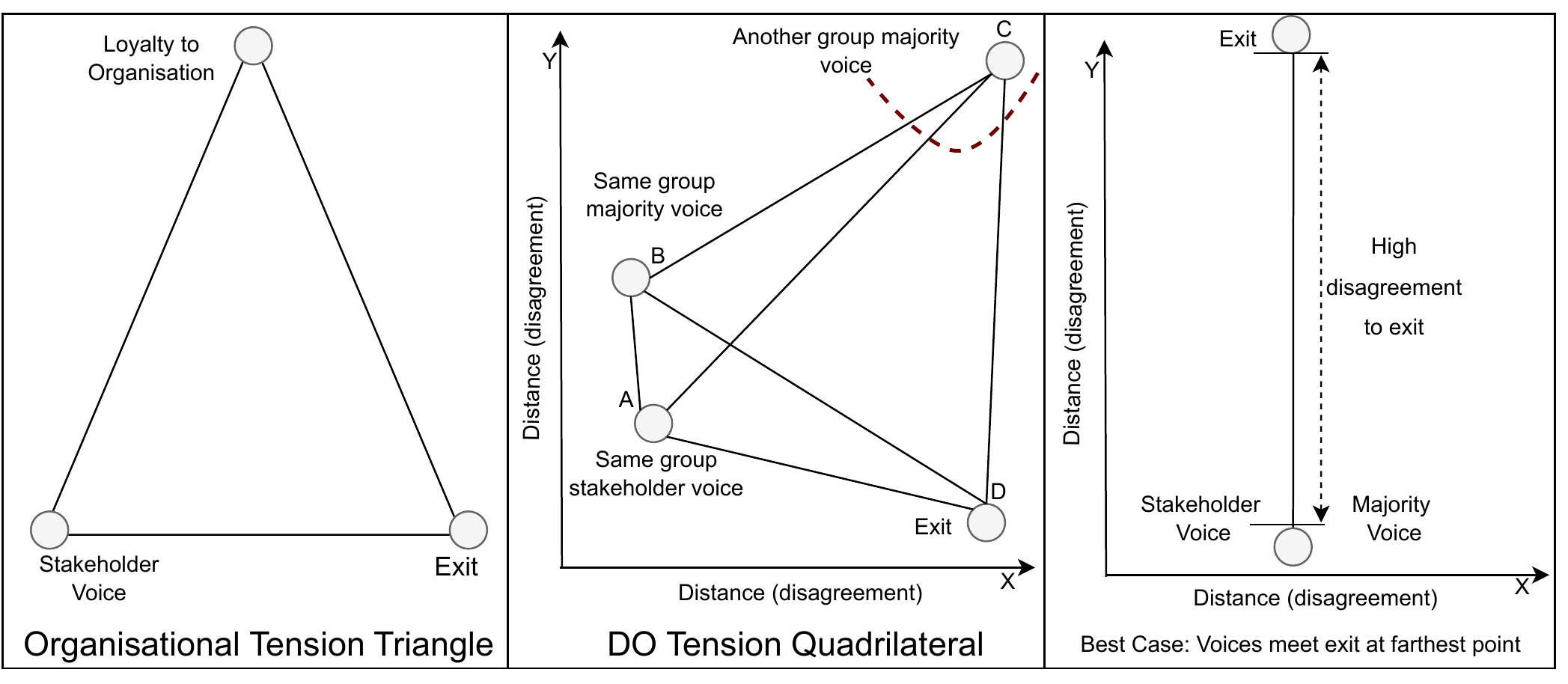}
	\caption{(a). A tension triangle portrays an unstable and mis-aligned set of views between a stakeholder and her organisation, with an option to exit. (b). When a DO consists of different groups, each with the autonomy to make relevant decisions — a tension quadrilateral (proposed) shows the tussle between a stakeholder and the different groups. Vertex A is the stakeholder voice   who belongs to the group whose majority voice is at vertex B. Vertex C is the majority voice of another group in the DO.  (c). The best case scenario is when stakeholder voice and majority voice are one and their distance to exit is maximised. }
	\label{fig:tensiontriangle}
\end{figure*}

\section{Background}
\label{sec:background}
\subsection{E-Voting}
\label{ssec:evoting}
Any e-voting approach, typically has the setup phase where the rules of the election are agreed upon, the
registration phase where voters are enrolled, the voting phase that involves the stakeholder submitting her vote choice to an electronic ballot box, followed by the tally phase which aggregates the vote choices and publishes the results.
The election discussed in this work is for 1-out-of-$y$ choices. I.e., the voter may vote for exactly 1 out of the $y$ available choices.
Recently, many blockchain-based e-voting approaches have been proposed~\cite{DBLP:conf/fc/SeifelnasrGY20,yu2018platform,killerprovotum,venugopalan2021bbbvoting}.  A blockchain enables not only to instantiate the immutable public bulletin board required for e-voting~\cite{Kiayias2002} but also provides censorship-resistance and correct execution of smart contract code, which are beneficial in this context, e.g., to verify the correctness of submitted votes and compute the tally.
Further, blockchains contribute to end-to-end verifiability~\cite{benaloh2015end,killerprovotum} and universal verifiability~\cite{Kiayias2002,yu2018platform}.

\subsection{Organisation Tension Triangle}
\label{ssec:Hirschman}
It highlights the following struggle to position oneself (a stakeholder) within an organisation  
(see Fig.~\ref{fig:tensiontriangle} (a).)
— should a stakeholder be loyal to the organisation and accept the policies imposed by the organisation despite her voice not being heard or exit the organisation?
According to Hirschman~\cite{Hirschman}, a stakeholder is able to choose from one of three options --- exit, loyalty or voice. With exit, a stakeholder  accepts to leave an unsatisfactory situation and changes her behaviour to get the best possible alternative result by exiting.
With loyalty, a stakeholder chooses to put up with new policies and not alter their behaviour, even when the policies were not agreeable to her.
With voice, a stakeholder makes her dissatisfaction with the new policy known to try and get the organisation or respective organisational groups (teams) to change or reverse it.

\subsection{Process Control Charts}
\label{ssec:pcrocesscharts}

A time series chart is a data visualisation chart that plots a series of data snapshots taken at regular intervals. The comparison provided makes it ideal to quickly identify  trends, spot outliers and to analyse how  key metrics change over time.
A process control chart is a time-series chart to monitor the acceptable limits of a particular process~\cite{Qui2013,Xie2002}. It uses real world  data to spot when a particular process is starting to move out of set control limits, so that its stakeholders may strategically intervene to resolve the issue.
When there are changes to a process, variations are natural and expected. Not all variations do  require intervention~\cite{Howard2003}. Setting control limits allows us to take note of the trends and prepare for intervention when it is necessary.  
Process control charts find application in a number of places such as industrial and manufacturing processes~\cite{Maia2012,Lefebvre2020}.

\section{System Model \& Design Goals}
\label{sec:modelandgoals}

\subsection{System Model}
\label{ssec:systemmodel}
The main actors are those involved in the running and participation   of the opinion poll.
Our model has the following main actors and components:
i) A stakeholder ($s\in S$)  casting a
vote for her choice ($C$) on a proposal $P$. ii) Election Authority ($EA$) who
is responsible for validating and registering stakeholders, and shifting between the
phases of the voting. iii) A blockchain smart contract ($SC$) collects the
votes, enforces
the rules of the voting and computes the tally of votes.
In Ethereum governance~\cite{ethgov2023},  stakeholders are classified as --- ETH holders, application users, application/tooling developers, node operators, EIP authors, validators and protocol developers.
%Also, some of the pitfalls in using coin based voting and a number of potential solutions are presented in Buterin~\cite{Buterin2021}.

\subsection{Need for repeated voting}
\label{ssec:needrepeatedvoting}
The use of repeated voting in blockchain decision making is not new. Bitcoin miners use it to arrive at a consensus, as part of the BIP-135\footnote{See https://en.bitcoin.it/wiki/BIP\_0135.}. Here, voting on miner support is carried out for up to 26 intervals (each interval comprising 2016 bitcoin blocks~\cite{Bhattacherjee2017}). In the case of BIP-135 these (among other) values may be configured.
In our opinion poll, we are required to capture the changing opinion of stakeholders over time.
%This may be a result of negotiations or due to other factors.
Stakeholder agreement may be subject to the incentives received via the tokens held, or reactions to external factors such as the narrative~\cite{Riley2015}, inflation~\cite{Shaun2009} and war~\cite{Verdickt2019}.
The agreeability of stakeholders with DO groups may also change with each new contentious proposal its  stakeholder(s) may not agree with.
To  capture time-varying information without the difficulties of restarting the process,  it will be run  repeatedly, and with minimal manual intervention.

\subsection{Design Goals}
\label{ssec:designgoals}

The following are the main design goals.
\begin{compactenum}
	
	\item Establish a generic method to enable the DO stability measurements (see Section~\ref{ssec:tensionquad}).
	
	\item Once initialised, the voting tool must be able to function mostly autonomous. I.e., continue to  repeatedly measure agreement between stakeholders and DO groups, except for some triggers that the $EA$ or any participating stakeholder may provide (see  Section~\ref{sec:repeatedvoting}.a).
	
	\item The voting tool must support both weighted  (staked) and equal weight voting (see Section~\ref{sec:votingtool}.c).
	
	\item The voting tool must be able to plug-in any suitable voting algorithm. The choice of the e-voting algorithm plugged-in to the voting tool is intentionally left to the implementer based on their exact requirements (see Section~\ref{sec:votingtool}.c).
	How stakeholders are assigned to group(s) and voting shares are supplied to them is left the DO. Its discussion is beyond the scope of this work.
	
	\item The voting tool  built must collect the agreement between the DO stakeholders and their own group/other groups in the organisation (see Section~\ref{sec:votingtool}).
	This  agreement must be quantified by providing a stability score (see Section~\ref{sec:experiments}).

\end{compactenum}

\section{Decentralised Organisation Stability}
\label{sec:dostability}

Stability may depend on how each stakeholder feels about the DO, and their interactions and relationship between the various groups in the organisation, at any given point in time.
In this section, we discuss the proposed DO tension quadrilateral used to enable stability measurements.
The organisational tension triangle introduced by Hirschman~\cite{Hirschman} and its effects on various organisational structures (see Section~\ref{ssec:Hirschman})  has been studied~\cite{Samman2020}. It highlights the  struggle to position oneself (a stakeholder) within an organisation.
The sides of the triangle (see Fig.~\ref{fig:tensiontriangle} (a.)) are not static and move around  based on the satisfaction of the stakeholder with the organisation.

\subsection{Tension Quadrilateral to enable stability measurements}
\label{ssec:tensionquad}

We draw from Hirschman's work to propose the DO tension quadrilateral.
In Fig.~\ref{fig:tensiontriangle} (b.), vertex A is the stakeholder voice  who belongs to the group whose majority voice is at vertex B. Vertex C is the majority voice of another group in the DO.
Any two groups may be at the same or different levels of the DO hierarchy.
Note, the hierarchies are not represented in Fig.~\ref{fig:tensiontriangle} (b.), or in the DO tension quadrilateral. For e.g., an upper hierarchical group need not necessarily appear above a lower hierarchical group. We are only interested in capturing distance (disagreement) between individual stakeholders and their own/different groups irrespective of the DO hierarchy and structure.

\begin{compactenum}
	
	\item [S1.] \textit{Measure individual stakeholder agreement with a different DO group}.  
	Close proximity of the vertices indicates agreement and distance is synonymous to disagreement. The stakeholder (in Fig.~\ref{fig:tensiontriangle} (b).) at vertex A is a member of the  group at vertex B.
	The distance between vertex A and  D is a measure of  stakeholder disagreement  with exiting  the DO. If these two vertices meet at the same point (0 distance),  the stakeholder is highly agreeable with exiting the DO. 
	The distance between vertex A and  C represents the  stakeholder disagreement with the majority voices of another group. 
	For example, in Ethereum it may be used to measure agreement of  an application developer (at vertex A), who is a member of the group at vertex B, with protocol developers who are members of group at vertex C  (see Section~\ref{ssec:systemmodel} for Ethereum stakeholders).
	
	\item [S2.] \textit{Measure  individual stakeholder agreement with own DO group}.  
	In this case, the stakeholder (in Fig.~\ref{fig:tensiontriangle} (b).) at vertex A is a member of the  group at vertex B. The distance between vertex A and  B represents the  stakeholder disagreement with the majority voices of her own group.
	%As  seen earlier, the distance between vertex A and D is a measure of the stakeholder disagreement with exiting the DO.
	For example, in Ethereum it may be used to measure agreement of a protocol developer at vertex A towards her group at vertex B by measuring the distance.
	
\end{compactenum}

These 2 cases encapsulates all the possibilities that may arise between stakeholders and DO groups.
In $S1$, if the majority of the same group stakeholders strongly disagree with another core group, then a majority of same group stakeholders might collectively exit the DO.
In $S2$, a stakeholder may exit if her agreeability to exit is far higher than the agreeability of aligning with the policy of majority stakeholders of her group.
The best case is when the agreeability of a stakeholder and the majority of stakeholders are  aligned to meet at a point, such that their exit distance is high (see Fig.~\ref{fig:tensiontriangle} (c.)). The worst case is when the  DO groups and exit meet at the same point. In practice, they lie on the vertices of a quadrilateral (or triangle), pulling or pushing at each other over time.
%For example, in Ethereum it may be used to measure agreement between (intra-group) protocol developers or  (inter-group) application and protocol developers (see Section~\ref{ssec:systemmodel} for Ethereum stakeholders). 

\section{Repeated Voting Framework}
\label{sec:repeatedvoting}
A stakeholder (voting participant) is permitted to vote repeatedly\footnote{Only contentious  proposals and those  deemed important for measuring DO stability are repeatedly voted (see Section~\ref{sec:intro} for strategic indicator sets).}, while the effect of her vote is manifested at the end of each voting interval, where votes are tallied and the results published (see Section~\ref{ssec:needrepeatedvoting} for justification in using repeated voting).
Our voting  repeats over a   fixed interval in the future.
To measure a quantity that varies over time, we use a repeated voting framework similar to Always-On-Voting~\cite{Venugopalan2021AoV} but with fixed  intervals.

\begin{figure}
	\centering
	\includegraphics[width=1.0\linewidth]{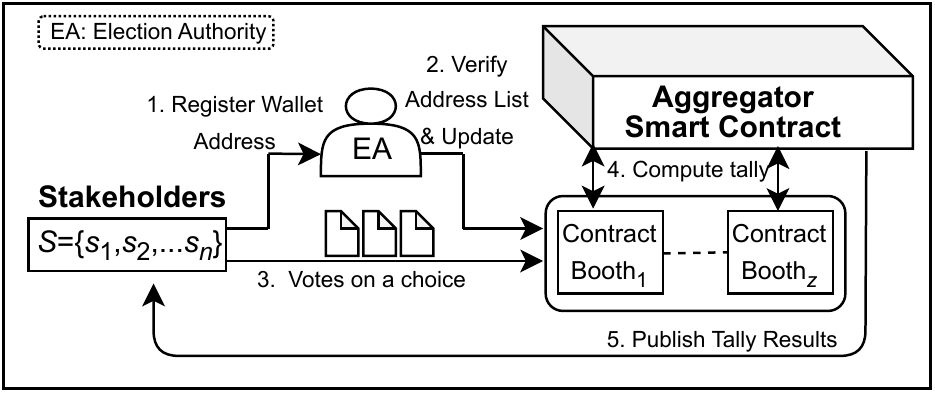}
	\caption{Interaction between stakeholders ($S$), election authority ($EA$),  contract booths and aggregator smart contract.
		(1) Registering wallet addresses of stakeholders and (2) their identity is verified by the $EA$ and the stakeholder list is updated on the smart contract.
		(3) Stakeholders send their vote to their assigned booth contract.	 
		The booth contract verifies the validity of the vote.
		(4) The aggregator contract is responsible for totaling individual booth tallies and (5) publicly announcing the total tally. }
	\label{fig:repeatedvoting}  
\end{figure}

The stakeholders (see Fig.~\ref{fig:repeatedvoting}),   register their wallet address with the $EA$ in step 1.
Without loss of generality, any method that can be used to verify identity would suffice.  In step 2, the $EA$ verifies and updates it on the booth smart contract\footnote{Participants are randomly grouped and assigned to booths  $\in\{1,2,...,z.\}$ (see Fig. \ref{fig:repeatedvoting}), represented by a booth smart contract. We use  multiple booth contracts to ensure its operations are modular.}.
The voting phase (in step 3), is where stakeholders cast their votes.
During tally computation (step 4), each booth contract $\{1,2,...,z\}$, computes its  local vote tally and sends it to the aggregator contract to find the total tally.
Next, the aggregator contract totals all the votes from each booth contract,  and publishes the final tally (step 5).
In repeated voting, the $EA$ is authorised to register/remove stakeholders in a future interval and update the stakeholder list on the smart contract.  
When there are no other changes in the next interval, revote repeats with step~3 and ends with step~5.

\begin{figure}
	\centering
	\includegraphics[width=1.0\linewidth]{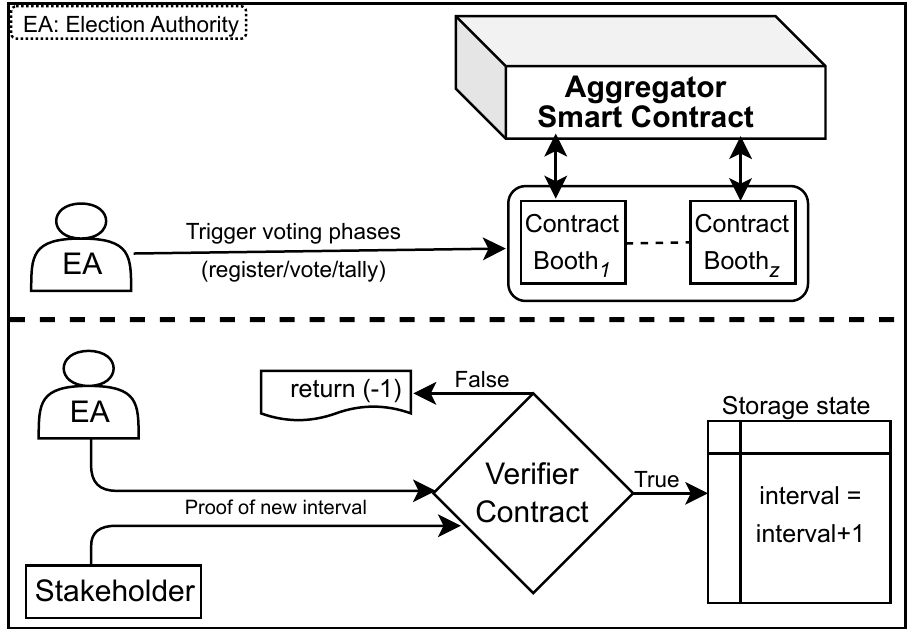}
	\caption{(Top). The $EA$ is responsible for shifting the phases of the election. Typically, the end of each phase marks the start of the next phase. The $EA$ is responsible for sending triggers to the booth contracts to move its state to the next phase. I.e., from registration to voting phase, and voting to tally phase.
		(Bottom). In repeated voting, the next election interval needs to be triggered after a required (fixed) period of time has elapsed. This trigger may be provided by either one of the stakeholders or the $EA$. This is achieved by supplying a proof to the verifier contract that the required period of time has elapsed. If the smart contract verification is successful, the internal storage state for the interval is incremented by 1.}
	\label{fig:trigger}  
\end{figure}

\paragraph{Triggering}
We employ two types of triggers to shift the state of the election (see Fig.~\ref{fig:trigger}).
The first type of trigger is to shift between the phases of the election (i.e., registration to voting phase and voting to tally phase). Typically, this trigger is provided by the $EA$ (see Fig.~\ref{fig:trigger}. (top)).
The second type of trigger is to start re-voting in the next interval (see Fig.~\ref{fig:trigger}. (bottom)).
This trigger may be supplied by the $EA$ or a voting participant.
However, it requires  a proof to be submitted to the verifier smart contract that a required amount of time has elapsed. If the verification passes, the internal state of the interval is incremented by 1 and the next election interval is initialised.
In some cases,  an explicit proof of elapsed time is not required to be provided by the $EA$ or a voting participant. When triggered, a smart contract is able to read the current block height ($h$) of the blockchain via a supported API call. This may be used to prove elapsed time. The rule of the smart contract can be set to increment the interval number by 1 (by comparing previous stored value of $h$ with its current value)  when at least X-blocks have been generated since the last trigger.
When there is no API support to call the current block height, we may use  multiple $EAs$ as oracles, each who certifies that a required interval time has elapsed. When at least $2/3^{rd}$ of the $EAs$ are in agreement and submit their certificate to the validator smart contract, it will trigger the next interval.

\paragraph{Cost of repeated voting}
The expenses from repeated voting on a public permissionless smart contract platform may be high. To reduce costs,
repeated voting can run on a public permissioned Proof-of-Authority
(PoA) blockchain, e.g., using Hyperledger projects such as
Besu.
Alternatively, smart contract platforms backed by trusted computing  may be used (e.g., Ekiden~\cite{cheng2019ekiden} and TeeChain~\cite{lind2019teechain}).
Here, expensive computations are moved to a trusted off chain  device.
Other partially-decentralised
layer 2 solutions such as Plasma, Polygon Matic, and Hydra  may also be used.
%The choice depends on the security and performance
Another option aimed at improving voting visibility and reducing overall costs is to allow block proposers (nodes) to signal support for the proposal.
Stakeholders may delegate their votes to the block proposer that supports her choice.
However, this requires the blockchain node code to be updated for it to be implemented.

\section{A Voting Tool to Collect Information}
\label{sec:votingtool}

\begin{figure*}
	\centering
	\includegraphics[width=1.0\linewidth]{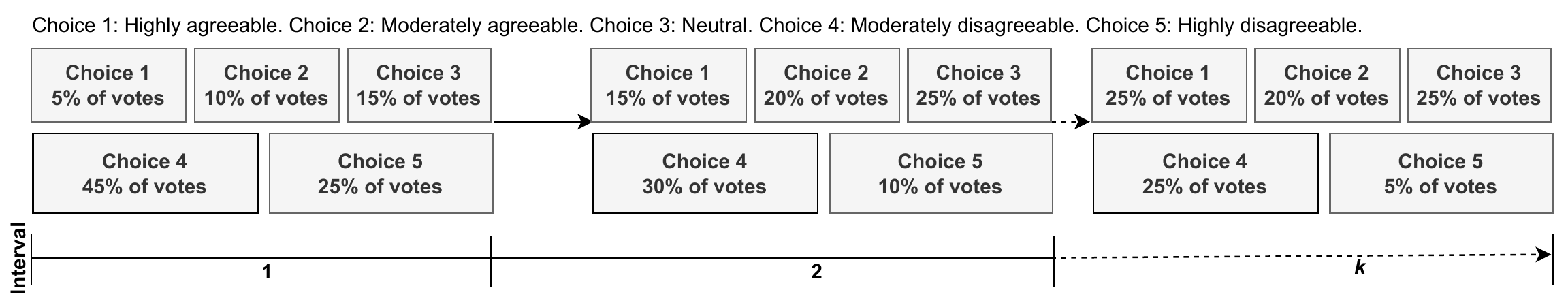}
	\caption{The elections repeat over fixed
		intervals (a.k.a., epochs). Here, the vote choices correspond to the agreeability of the stakeholder to the proposal being voted. The choices are arranged in  a linearly increasing scale, where choice 1 is the most agreeable and choice 5 is the least agreeable.
		Next, repeated voting within $k$ fixed intervals results in a choice transition towards highly agreeable.
		The tally is computed at the end of each interval.}
	\label{fig:transition-epoch}  
\end{figure*}

Opinion polls are regularly used in market research and to gauge support in public policy decision making~\cite{monero2019}.
In this section, we combine the ideas presented in DO tension quadrilateral (Section~\ref{ssec:tensionquad}) and repeated voting framework (Section~\ref{sec:repeatedvoting}), to collect information from the DO stakeholders by running an opinion poll.
Any suitable voting algorithm may be plugged-in to the repeated voting framework to collect information. Note that our opinion poll only changes how the vote choices are encoded for the purpose of collecting  information and repeating the election over fixed  intervals.
In our voting tool, we encode vote choices as a measure of agreement the individual (stakeholder) has towards DO groups and DO exit.
It may also be used to encode stakeholder agreement towards contentious proposals.
Any DO stakeholder  may  vote their choice in any new voting interval to show their agreement/disagreement.
The votes are tallied at the end of each interval.

\paragraph{Encoding vote choices} A 1-out-of-$y$ voting in a DO is defined by the 4 tuple ($ S, P, C, T $). The stakeholders ($S$) may vote on any proposal ($P$) for any 1 of their $y$ vote choices ($c_1,c_2,\ldots,c_y\in C$).
A tally ($T$) computes the sum of votes received for each choice  in $C$.
For each proposal, we provide 5 vote choices ($C=c_1,c_2,\ldots,c_5$). The choices represent a linear scale arranged in ascending order of disagreement (see Section~\ref{ssec:tensionquad}). I.e., ($c_1 \rightarrow 0,\ldots,c_5 \rightarrow 4$).
For the DO quadrilateral (see Fig~\ref{fig:tensiontriangle} (b.)), we may have  a proposal  $P_1$  asking its same group stakeholders, ``what is your agreement w.r.t to your own group on a scale of 0 to 4? (0  highly agreeable and 4 highly disagreeable)''. I.e., provide the distance between vertex A and B in Fig~\ref{fig:tensiontriangle} (b.).
If her vote choice is 0, her disagreement with her group is 0 (none). Hence, the distance to her group is 0. Both vertices A and B will meet at the same point. This is the best case individual outcome for this proposal. If her vote choice is 4, she is in high disagreement with her group. Vertices A and B will be at the farthest distance from each other.
Another proposal $P_2$ may ask a stakeholder, ``what is your agreement with another specified core  group in your organisation on a scale of 0 to 4? (0  highly agreeable and 4 highly disagreeable)''. I.e., provide the distance between vertex A and C in Fig~\ref{fig:tensiontriangle} (b.).
A third proposal $P_3$  may ask ``what is your agreement as a  stakeholder  to exiting the DO? (0  highly agreeable to exit and 4 highly disagreeable)''. I.e., provide the distance between vertex A and D in Fig~\ref{fig:tensiontriangle} (b.).
A vote choice of 0 implies the stakeholder is in high agreement with the proposal of exiting the DO and a vote choice of 4 implies high disagreement with exiting the DO.

\paragraph{Repeated voting for collecting time series data} Repeated voting  is incorporated to repeatedly input  stakeholder agreement  (over time) to our voting tool.
For the example given in Fig.~\ref{fig:transition-epoch}, there are 5 choices. Each choice represents the level of agreeability to the proposal. Choice 1 denotes the highest level of agreeability and choice 5 represents the least agreeability. At the end of the interval 1, choices 1 and 2 combined, constitute only 15\% of the votes. When choice 3 is taken as the midpoint, we observe there are more votes towards disagreeability.
However, over repeated voting intervals $1, 2, ..., k$, the votes are seen to shift towards agreeability. At the end of  interval $k$, choices 1 and 2 combined tallies to 45\% of the votes, indicating a stronger level of agreeability.

\paragraph{Algorithm for collecting agreeability information}
Our algorithm also supports weighted voting. I.e., it is suitable for instances where some stakeholders have a higher voting share (to indicate power holder agreement) when compared to others.
Let $n$ be the total number of stakeholders.
For weighted voting, when a stakeholder $s_{j\in n}$ has a voting power of $w_j\in \mathbb{Z}^+$ (a positive integer) votes, it is counted as $w_j$ votes for the choice voted.
Alternatively, it may also be used when all stakeholders have equal voting power, i.e., $w_j=1,\forall j\in n$.
The main steps used in repeated voting to collect stakeholder agreement are shown in Algorithm. \ref{alg:tensionquad}. It involves 3 main functions.
The function $VoteInInterval$ allows an individual stakeholder in the current voting interval to vote  on proposal $P_x$, for her desired choice (lines 12-15).
A proof $\pi_j$ is used to prove voting eligibility  of the $j^{th}$ stakeholder\footnote{For e.g., a proof of digital identity verified against registered voters.}.
However, when the current voting interval ends, and the interval is updated (to the next interval), votes to previous intervals can no longer be added or changed.  The function $TallyInInterval$ is called to tally the votes in any given interval. It returns a tally as an indicator of aggregate agreement (line 24), for each of the 5 voting choices.
The  function $Updateinterval$ (lines 25-29) is used to update the election to its next voting interval.
A proof $\pi_{nxtint}$ is used to trigger the next voting interval\footnote{For e.g., a proof the blockchain mined X-blocks since the last trigger.}. These proofs are verified for its correctness before taking the appropriate action.

All steps, including the verification of proofs are carried out using smart contracts.
The election authority ($EA$) is responsible for updating the smart contract with the latest list of registered stakeholders.
Any stakeholder or $EA$ with a valid proof $\pi_{nxtint}$, may trigger the next voting interval (see Section~\ref{sec:repeatedvoting} for trigger mechanisms).
As discussed in Section~\ref{ssec:evoting}, a number of blockchain e-voting protocols are available to be plugged-in to the repeated voting framework.
Based on the exact requirements, the e-voting protocol used is left to the implementer.

\begin{algorithm}[t]
	\label{alg:tensionquad}
	\DontPrintSemicolon
	
	\footnotesize
	\SetKwProg{Fn}{Function}{:}{\KwRet}
	\SetKwFunction{VoteInInterval}{VoteInInterval}
	\SetKwFunction{TallyInInterval}{TallyInInterval}
	\SetKwFunction{UpdateInterval}{UpdateInterval}
	
	Let stakeholders $S=\{s_1,\ldots ,s_n\}$.\\
	Let $\pi_j$ be proof of identity for $j^{th}$ stakeholder, $1\leq j \leq n$.\\
	Let $\pi_{nxtint}$ be proof the trigger to update voting interval is valid.\\
	Let $w_j$ be voting shares held by stakeholder $s_j$.\\
	
	Let $P_x$ be the $x^{th}$ DO proposal.\\
	Let choices $C=\{c_1,\ldots ,c_{y=5}\}$.\\
	Let $c\in C$ be the vote choice of $s_j$ in voting interval $i$ for $P_x$.\\
	Let $T=\{T_1,\ldots ,T_{5}\}$ be the interval tally for each choice.
	\\Initialise $T \leftarrow 0$,$interval\leftarrow 0$, $vote\leftarrow 0$.\\
	
	Input. $P_x, S, C, T, interval,\pi_j$,$\pi_{nxtint}$. \\
	Output. Stability measure ($T_1,\ldots,T_{5}$)$\in T$,$\forall$ ($c_1,\ldots,c_{y=5}$)$\in P_x$ in each  voting interval $i$.
	
	% Write Function with word ``Def''
	\SetKwProg{Fn}{Def}{:}{}
	\Fn{\VoteInInterval{$x, interval, j, c, v_{ij}, \pi_j$}}{    
		$i \leftarrow interval$\\
		\If{($isIntervalCurrent(i)$ AND $ verifyVoter(j,\pi_j)$)}
		{
			$vote[x][i][j][c]\leftarrow  w_{j}$\tcp{Add to storage}
		}
	}

	\Fn{\TallyInInterval{$x, interval,vote$}}{
		$T\gets 0$\\    
		\If{($isIntervalExisting(interval)==false$)}
		{return -1 \tcp{return failed check}}
		\For{($choice\gets$ 1 to $y$) }
		{
			\For{($stakeholder\gets$ 1 to $n$) }
			{
				\If{$vote[x][interval][stakeholder][choice]>0$}    
				{
					\tcp{Compute tally for each choice}
					$T[x][interval][choice] += vote[x][interval][stakeholder][choice]$\\
					
				}
			}
		}
		return $T$
		
	}
	
	\Fn{\UpdateInterval{$nxtint,\pi_{nxtint}$}}{    
		\If{($IntervalUpdateVerify(nxtint,\pi_{nxtint})$)}
		{
			$k\leftarrow  readState(interval)$\tcp{Read  from storage}
			$interval\leftarrow  k+1$\\
			$writeState(interval)$ \tcp{Write to storage}    
		}
	}
	
	\caption{Voting tool to collect  information}
\end{algorithm}

\section{Experiments \& Measurements}
\label{sec:experiments}

\begin{figure*}
	\centering
	\begin{minipage}[b]{.45\textwidth}
		\includegraphics[width=1.0\linewidth]{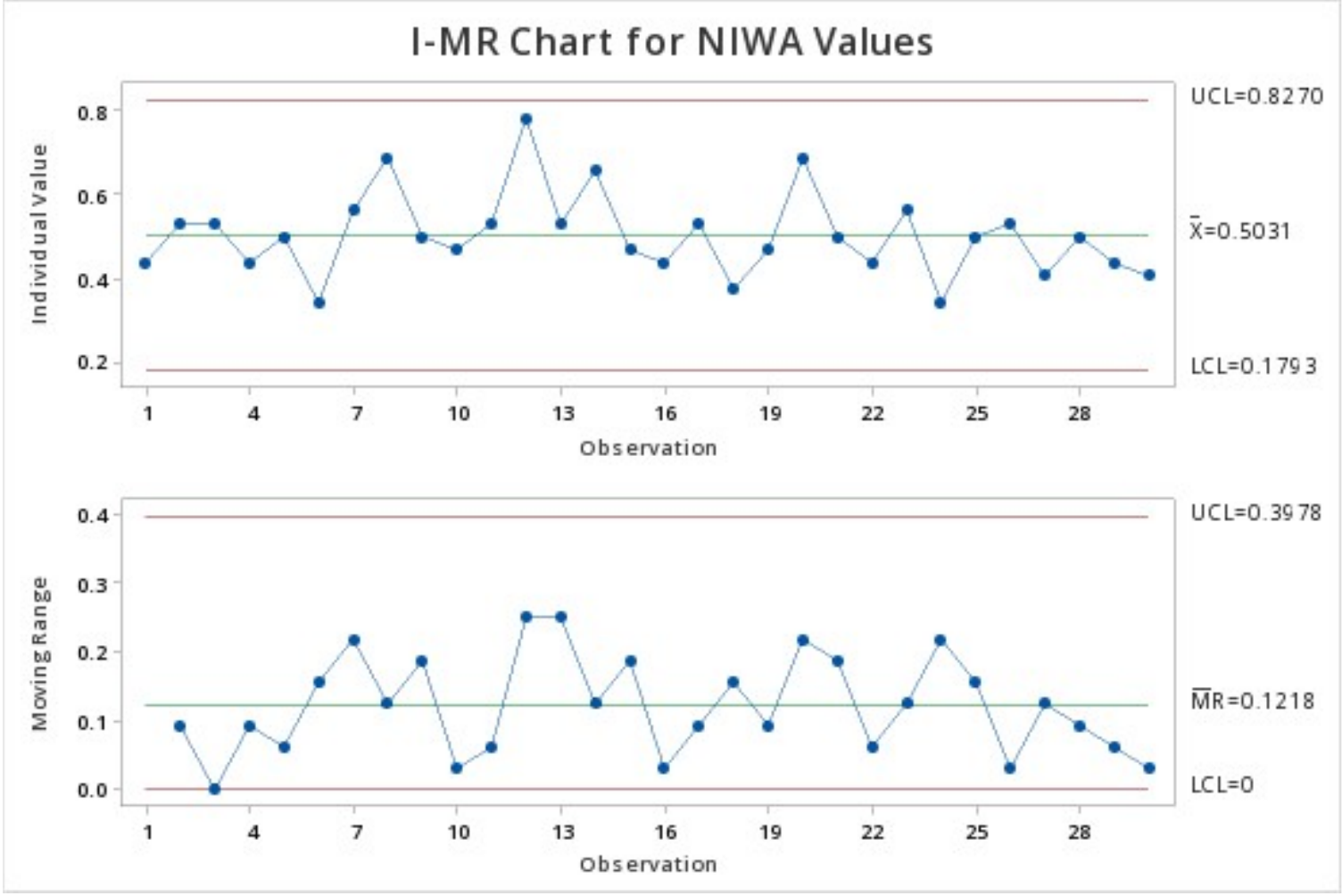}
		\caption{Random sampling: I-chart (top) displays the individual $NIWA$ values for the observation/intervals and  MR chart (bottom) displays the moving range. The two charts are viewed in tandem to detect individual and range variations. All data points in the experiment are seen to be within the control limits. }\label{fig:imr}
	\end{minipage}\qquad
	\begin{minipage}[b]{.45\textwidth}
		\includegraphics[width=1.0\linewidth]{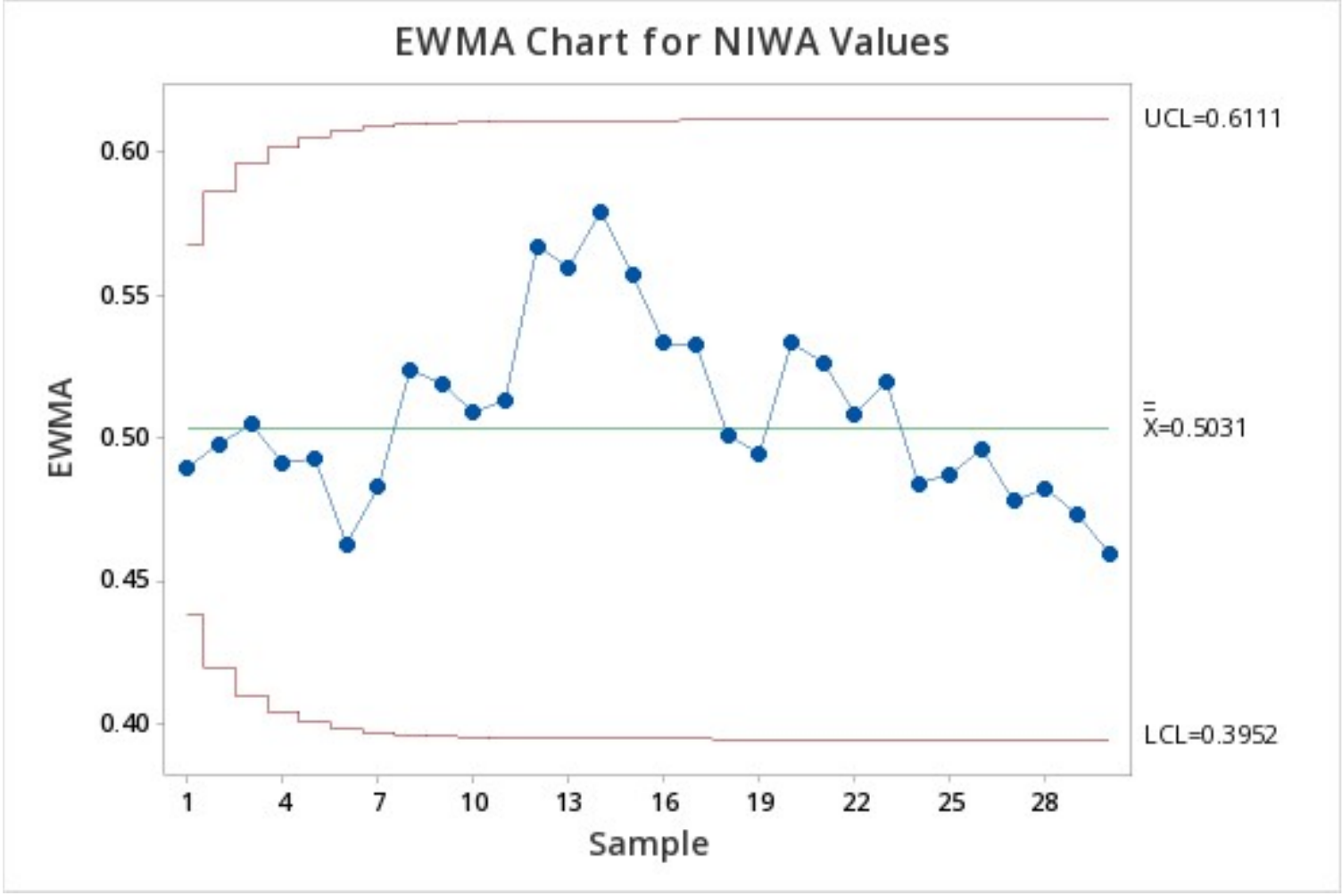}
		\caption{Random sampling: The EWMA chart is seen to be sensitive to recent data points since they are assigned a higher weightage. All data points in the experiment are seen to be within its control limits.
		}\label{fig:ewma}
	\end{minipage}
\end{figure*}

\begin{figure*}
	\centering
	\begin{minipage}[b]{.45\textwidth}
		\includegraphics[width=1.0\linewidth]{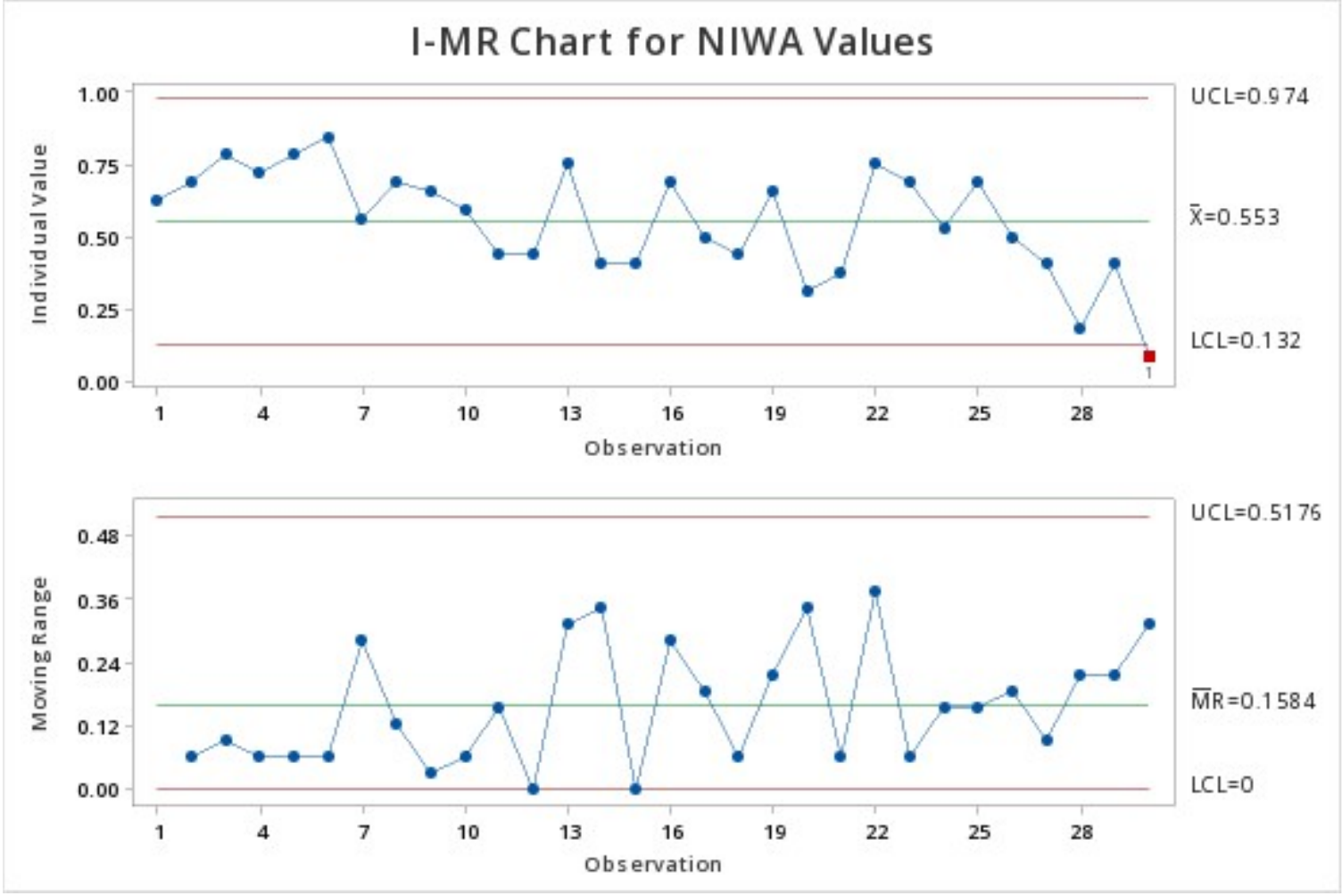}
		\caption{Purposive sampling: I-chart (top) displays the individual $NIWA$ values for the observation/intervals. It  is beyond the control limit at interval 30. MR chart (bottom) displays its moving range.}\label{fig:imr2}
	\end{minipage}\qquad
	\begin{minipage}[b]{.45\textwidth}
		\includegraphics[width=1.0\linewidth]{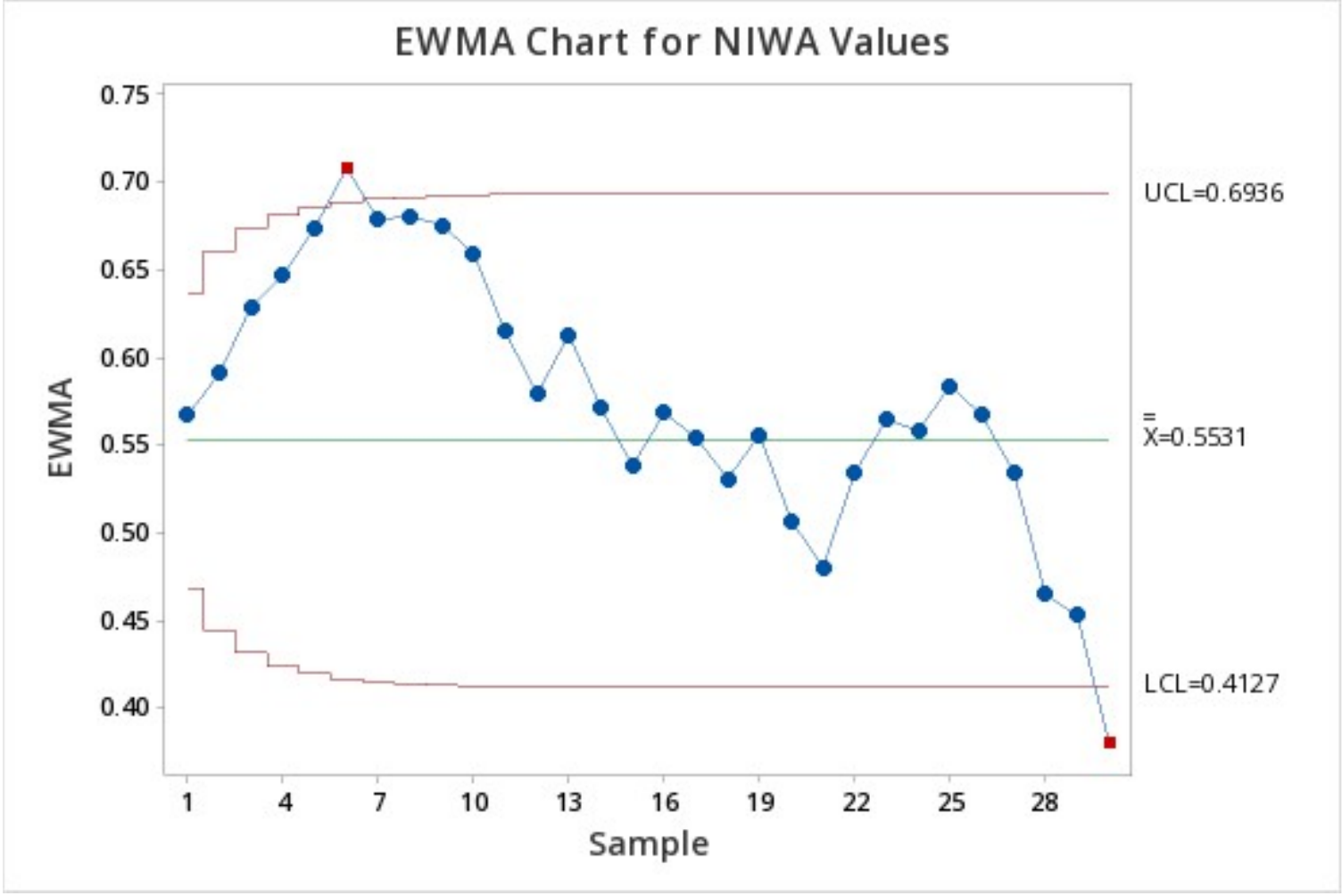}
		\caption{Purposive sampling: The EWMA chart for the experiment is seen to be out of the control limits at  intervals 6 and 30. }\label{fig:ewma2}
	\end{minipage}
\end{figure*}

Once the agreeability information from the opinion poll is collected as vote tallies (see Algorithm~\ref{alg:tensionquad}), the DO stability measurements may  be carried out.
Five vote choices are provided in Fig.~\ref{fig:transition-epoch}.
The  choices are --- $c_1$ (highly agreeable), $c_2$ (moderately agreeable), $c_3$ (neutral), $c_4$ (moderately disagreeable) and $c_5$ (highly disagreeable).
The votes in any given interval are tallied using Algorithm.~\ref{alg:tensionquad}. Its output is
($T_1,\ldots,T_{5}$)$\in T$,$\forall$ ($c_1,\ldots,c_{y=5}$)$\in P_x$. I.e., summation of votes (tallies) for each choice, given a  proposal $x$ and a  voting interval.
In this section, we convert the interval tallies to  a measurable score representative of   agreement to a proposal or DO stability for the interval and explain its working. Further, these scores are plotted across intervals on time series charts to study its key indicator trends over time.

We assign weights for each of the 5 vote choices.
Choice 3  (neutral choice in Fig.~\ref{fig:transition-epoch}) is assigned  as the midpoint. Positive weights appear to the  left of choice 3 and negative weights appear to its right. The choices farther away from the midpoint are assigned  incrementally higher absolute weights.
Therefore, choice 1 is assigned a weight 2/5, choice 2 is assigned  1/5, choice 3 is assigned 0/5, choice 4 is assigned -1/5 and choice 5 is assigned -2/5.
Let there be a total of $T$ votes, where $T_1$ is total votes received for choice 1, $T_2$ for choice 2, $T_3$   for choice 3, $T_4$   for choice 4 and $T_5$  for choice 5, such that
\begin{equation}
	\label{eqn_1}
	T = T_1 + T_2 + T_3 + T_4 + T_5
\end{equation}

Next, the normalised weighted average  is computed as
\begin{equation}
	\label{eqn_2}
	NWA = \frac{\left( \frac{2}{5} \cdot T_1+ \frac{1}{5} \cdot T_2+ \frac{0}{5} \cdot T_3+ \frac{-1}{5} \cdot T_4+ \frac{-2}{5} \cdot T_5\right)\cdot \frac{5}{2}}{T}
\end{equation}
and the normalised interval weighted average (NIWA) is calculated as
\begin{equation}\label{eqn_3}
	NIWA = \left(1-  \frac{NWA+1}{2}\right)
\end{equation}

The numerator in Equation.~\ref{eqn_2} is multiplied  with 5/2
to normalise the weighted average into the interval [-1, 1].
The value of $NWA$ in Equation.~\ref{eqn_3} is mapped into the interval [0, 1] by computing $1-\frac{NWA+1}{2}$.
A $NIWA$ score towards 0 indicates agreeability and a score towards  1 indicates disagreeability to the proposal.
A midpoint score of 0.5 is a neutral score.

Note that each NIWA score is computed for a single voting interval, given a proposal.  
This score provides an agreement measure (value) for the  interval but it does not capture the change in trends over time.
To achieve this, we use the time series data from repeated voting intervals to capture the  changes and to check whether the process is stable or not (see Section~\ref{sec:background}.C. for background).

To capture the change in trends, we plot an Individual-Moving Range (I-MR) chart~\cite{minitabimr2013} and an Exponential Weighted Moving Average (EWMA) chart~\cite{Padilla2020}.
An I-MR chart consists of two charts --- an Individual (I) and Moving Range (MR) chart. The I-chart plots individual (NIWA) data points over a specified set of ordered intervals and a MR chart plots its moving range.
Its mean is denoted as $\bar{x}$ and the standard deviation is $\sigma$. The lower control limit (LCL) is marked at $\bar{x}-3\cdot \sigma$ and the upper control limit (UCL) is at  $\bar{x}+3\cdot \sigma$.

\paragraph{Random sampling experiment} We used synthetic data as input to derive our  measurements. For each voter, her vote was chosen uniformly at random from the 5 vote (agreeability) choices.
We used Python 3.x to generate pseudo-random numbers, generate votes, tally votes in an interval, and to program the logic for Equation.~\ref{eqn_1}-\ref{eqn_3}.
The experiment  included voting in  30 consecutive intervals, i.e., 30 observations  were made. A total of 8 stakeholders (voters) cast their votes.
The 30 NIWA values were computed using Equation~\ref{eqn_3} and stored in a file.
We used the  Minitab statistical software~\cite{Minitab2023}  to generate the I-MR chart (Fig.~\ref{fig:imr}) and EWMA chart (Fig.~\ref{fig:ewma}).
The NIWA scores stored in the file were the input required to generate the time series charts.
For the I-chart (see top chart in Fig.~\ref{fig:imr}), the ordered intervals are plotted on the x-axis  and the  corresponding $NIWA$ values are plotted on the y-axis.
All $NIWA$ values for this experiment are seen to be within 3 standard deviations from the mean, Hence, the process is stable\footnote{According to Chebyshev's inequality~\cite{Frost2021}, if the process is stable, 89\% of the time the data point will fall within $\bar{x}\pm 3\cdot \sigma$ , irrespective of the form of the distribution~\cite{Howard2003}.
}.
The moving range is the difference between two successive data points in the I-chart. The MR chart (see bottom chart in Fig.~\ref{fig:imr}) shows the variability of the range.
For the EWMA chart, each previous mean sample is assigned a weight. The most recent values are weighted the highest and the oldest values are weighted the least.
The values are weighed in geometrically decreasing order.
The EWMA chart is sensitive to small shifts in the process mean, whereas the I-MR chart is sensitive to larger shifts. Hence, both charts are used side-by-side to improve detection.
The EWMA chart in Fig.~\ref{fig:ewma} used Minitab statistical software and weight $\lambda=0.2$.
The EWMA chart shows a decreasing trend and its most recent data points are seen to be below the mean.

\subsection{Purposive sampling experiment}
\label{ssec:Purposivesampling}

In this experiment, the data points are sampled closely to a real world scenario.
To generate a non-uniform random sample, we used the Python NumPy API call
\begin{lstlisting}
	np.random.choice([c1,c2,c3,c4,c5],
	p=[p1, p2, p3, p4, p5])
\end{lstlisting}
It is used to pick 1-out-of-5 vote choices based on the respective weighted probabilities in p. Note that p = p1 + p2 + p3 + p4 + p5  and p = 1.
Initially, we picked a higher probability for generating choice 4 (c4) and choice 5 (c5) by setting a higher value for p4 and p5, respectively.
For each successive interval, we incremented the probability p1   and decremented the remaining probabilities p2 to p5 by a constant step size.
As a result, the voting probability  for the first choice steadily increased and the voting probabilities  for the remaining choices  decreased.
A total of 8 voters cast their vote in our experiment.
As seen earlier, the vote tallies were computed using Algorithm.~\ref{alg:tensionquad}.
It was run through Equation.~\ref{eqn_1}-\ref{eqn_3} to find the NIWA score for the interval.
The data points (NIWA values) were collected for 30 ordered intervals.
The I-chart for this experiment (see Fig.~\ref{fig:imr2}, top chart) was seen to have 1 data point beyond the control limit. The NIWA value at interval (observation) 30 is three  standard deviations below the mean.
The EWMA chart for this experiment (see Fig.~\ref{fig:ewma2}) had 2 data points beyond the control limits. The one at interval (sample) 6 is  at three  standard deviations above the mean  and the other at interval 30 is three  standard deviations below the mean.
When the data points are beyond their control limits, action may  be taken to rectify the situation and bring the process back within the control limits (for an unfavourable outcome).
For e.g., it may be worth investigating the I-chart in Fig.~\ref{fig:imr2}. Here, the NIWA value at interval 30 is below LCL and tending towards 0. If this chart were for the proposal --- “what is the polled stakeholders agreeability with exiting the organisation (on a scale of 0 to 4)?”, a NIWA score towards 0 would imply  a high agreement towards exiting the DO.

\section{Conclusions}
\label{sec:conclusions}
We modified the Hirschman's tension triangle and adapted it to a DO, to create the DO tension quadrilateral. It  was used to enable  DO stability measurements.
We proposed a voting tool by encoding stakeholder agreement onto vote choices. This tool aggregated stakeholder agreeability information as vote tallies.
The vote tallies were used as the input to develop a  stability score. The stability scores over consecutive intervals  were plotted as time series charts to observe its key indicator trends.
The stability measurements may be used to find common ground between the stakeholders and the DO. The  chart  trends act as an early warning system for DO destabilisation and exit.

\section*{Acknowledgment}
This research is supported by the National Research Foundation, under its Campus for Research Excellence and Technological Enterprise (CREATE) Programme.

\bibliographystyle{IEEEtran}
\bibliography{ref-reduced}
\end{document}